
\magnification=1200
\baselineskip=13pt
\overfullrule=0pt
\tolerance=100000

{\hfill \hbox{\vbox{\settabs 1\columns
\+ UR-1416 \cr
\+ ER-40685-866\cr
\+ hep-th/9504059\cr
}}}

\bigskip

\baselineskip=18pt

\centerline{\bf DYNAMICAL SUPERSYMMETRY}

\bigskip
\bigskip

\centerline{Ashok Das}
\centerline{and}
\centerline{Marcelo Hott$^\dagger$}
\centerline{Department of Physics and Astronomy}
\centerline{University of Rochester}
\centerline{Rochester, NY 14627}

\bigskip
\bigskip
\bigskip
\bigskip

\centerline{\bf Abstract}

\medskip
We show, in a simple quantum mechanical model, how a theory can become
supersymmetric in the presence of interactions even when  the
free theory is not. This dynamical generation of supersymmetry
relaxes the condition on the equality of masses of the
superpartners which would be of phenomenological interest.

\vskip 2 truein

\noindent $^\dagger$On leave
 of absence from UNESP - Campus de
Guaratinguet\'a, P.O. Box 205, CEP : 12.500, Guaratinguet\'a, S.P., Brazil

\vfill\eject

Supersymmetry [1] is a rich theoretical concept which relates bosons to
fermions.
It has found applications in many diverse areas of physics [2-4]. It is
also of great
interest in phenomenological studies in high energy physics [5].
Conventionally, in
the study of supersymmetric theories, one starts with a free theory which is
invariant under supersymmetry transformations (transformations which take
bosons
into fermions and vice versa). This requires bosons and fermions (the
superpartners) of the theory to have equal masses (or frequencies if one is
dealing with quantum mechanical oscillators). Interactions are then introduced
so as to maintain the tree level supersymmetry or build on it. Namely, the
supersymmetry transformations of the interacting theory (if they are different
from the tree level transformations) reduce to the tree level ones when
interactions are switched off.

Supersymmetric theories have many interesting properties and that is, of
course, the main reason for all the interest in such theories. However, the
equality of masses for the superpartners is a worriesome feature of these
theories since bosons and fermions with degenerate masses are not observed in
nature. Spontaneous breaking of supersymmetry, on the other hand, is
technically nontrivial compared to the breaking of ordinary symmetries which
complicates the study of supersymmetric phenomenology. It will, therefore, be
of great help if, somehow, the condition of equality of masses can  be relaxed
in supersymmetric theories. In this letter, we will show within the context of
a quantum mechanical model how this can be achieved. More specifically, we will
start with a free theory of a bosonic and a fermionic oscillator of unequal
frequencies which is not supersymmetric and show that in the presence of
interactions this theory can become supersymmetric. This is what we call
dynamical supersymmetry and it does not require the boson and the fermion to
have equal frequencies (masses).

Let us start with a quantum mechanical theory of a free bosonic and fermionic
oscillator described by the Hamiltonian,

$$  H_0 = \omega a^\dagger a + \epsilon c^\dagger c\eqno(1)$$
where $a$ and $c$ stand for the bosonic and the fermionic annihilation
operator respectively with $\omega$ and $\epsilon$ representing their
respective frequencies. The creation and the annihilation operators for the
bosons (fermions) satisfy the standard (anti) commutation relations[6].
As is well known, when $\omega=\epsilon$, this defines
the supersymmetric oscillator [6-8] which is invariant under the supersymmetric
transformations generated by the supercharges [6]

$$  Q = a^\dagger c\,\,\,\,\, {\rm and}\,\,\,\, \overline Q = c^\dagger a
\eqno(2)$$

In our entire discussion, however, we will assume that $\omega\neq \epsilon$.
Our starting theory is, therefore, not supersymmetric since the bosonic and the
fermionic frequencies (masses) are not equal. However, let us now
look at the following interacting Hamiltonian [9-10],

$$ H =  \omega a^\dagger a + \epsilon c^\dagger c + g(a^\dagger+a)c^\dagger c
\eqno(3)$$
where $g$ represents the strength of the interaction. We will now show that for
the specific value of the coupling parameter (We assume $\epsilon >\omega$.)
$$  \epsilon - \omega = {g^2\over\omega}\eqno(4)$$
the Hamiltonian in Eq.(3) becomes supersymmetric.

To show this, let us consider the fermionic charges
$$\eqalign{ Q & = a^\dagger c\exp{({g\over\omega}(a^\dagger - a))}\cr
\overline Q & = \exp{(-{g\over\omega}(a^\dagger - a))} c^\dagger a}\eqno(5)$$
With the standard (anti) commutation relations of the theory, it is
straightforward to show that
$$\eqalign{[Q,H] & = (\epsilon -\omega -{g^2\over\omega}) Q\cr
[\overline Q,H] & = -(\epsilon -\omega -{g^2\over\omega}) \overline Q}
\eqno(6)$$
It is clear now that when the condition in Eq.(4) holds, these fermionic
charges are conserved and define supersymmetric transformations under which the
interacting Hamiltonian in Eq.(3) is invariant. It is also straightforward to
show that
$$[Q,\overline Q]_+ = {1\over\omega}[ H -
(\epsilon-\omega-{g^2\over\omega})c^\dagger c ]\eqno(7)$$
This shows that the conserved charges $Q$ and $\overline Q$ satisfy the
conventional supersymmetry algebra when Eq.(4) holds. The Hamiltonian $H$ of
Eq.(3) can be easily checked (with the condition in Eq.(4))
to be invariant under the supersymmetry transformations

$$\eqalign{\delta a & =
-\lambda(1+{g\over\omega}a^\dagger)\exp{({g\over\omega}(a^\dagger - a))}c\cr
\delta a^\dagger & = -{g\over\omega}\lambda a^\dagger
c\exp{({g\over\omega}(a^\dagger - a))}\cr
\delta c & = 0\cr
\delta c^\dagger & = \lambda a^\dagger\exp{({g\over\omega}(a^\dagger - a))}
}\eqno(8)$$
and
$$\eqalign{\overline\delta a & =
{g\over\omega}\overline\lambda\exp{(-{g\over\omega}(a^\dagger - a))}c^\dagger
a\cr
\overline\delta a^\dagger & = \overline\lambda\exp{(-{g\over\omega}(a^\dagger -
a))}c^\dagger (1 + {g\over\omega} a)\cr
\overline\delta c & = \overline\lambda\exp{(-{g\over\omega}(a^\dagger - a))}
a\cr
\overline\delta c^\dagger & = 0
}\eqno(9)$$
Here $\lambda$ and $\overline\lambda$ are the two constant Grassmann parameters
of the supersymmetry transformations.

Thus, we see that even though the starting theory is not supersymmetric and the
bosonic and the fermionic oscillators have different frequencies (masses), for
a particular value of the interaction strength, the interacting Hamiltonian has
become supersymmetric. The theory has generated supersymmetry dynamically.
Since the bosons and the fermions correspond to different frequencies, it is
worth investigating the structure of the supersymmetric spectrum of states in
this theory. It can be easily checked that the superpartner states now involve
coherent states in a nontrivial way ( Eq.(4) is assumed.).

$$\eqalign{ Q|n_a,n_c=1\rangle & ={1\over\sqrt{n_a!}} a^\dagger(a^\dagger -
{g\over\omega})^{n_a}|{g\over\omega},n_c=0\rangle\cr
\overline Q|n_a+1,n_c=0\rangle & = \sqrt{{n_a+1\over n_a!}}(a^\dagger +
{g\over\omega})^{n_a}|-{g\over\omega},n_c=1\rangle
}\eqno(10)$$
Here we have introduced the coherent states defined by [11]

$$|\alpha,n_c\rangle = \exp{(\alpha(a^\dagger -
a))}|n_a=0,n_c\rangle\eqno(11)$$
 Thus, we see that the relation between the
perturbative supersymmetric partner states, in
this case, are not as simple as in the conventional supersymmetric theories.

Finally, let us note here that this theory can be exactly solved and all of the
above features can be seen in a simpler way as follows. Let us define a
generalized Bogoliubov transformation defined by the operator
$$ U = \exp{(-{g\over\omega}(a^\dagger - a)c^\dagger c)}\eqno(12)$$
This defines a unitary transformation leading to

$$\eqalign{ b & = UaU^\dagger = (a + {g\over\omega}c^\dagger c)\cr
b^\dagger & = Ua^\dagger U^\dagger = (a^\dagger + {g\over\omega}c^\dagger c)\cr
f & = UcU^\dagger = \exp{({g\over\omega}(a^\dagger - a))}c\cr
f^\dagger & = c^\dagger\exp{(-{g\over\omega}(a^\dagger - a))}
}\eqno(13)$$

These new variables satisfy the canonical (anti) commutation relations like the
original fields since the transformation is unitary. It is now straightforward
to check that the interacting Hamiltonian of Eq.(3) can be rewritten in terms
of these new variables as
$$ H = \omega b^\dagger b + (\epsilon - {g^2\over\omega})f^\dagger f\eqno(14)$$
The energy eigenstates and the eigenvalues in terms of these variables are
quite simple, namely,

$$ H|n_b,n_f\rangle = E_{n_b,n_f}|n_b,n_f\rangle\eqno(15)$$
with
$$ E_{n_b,n_f} = \omega n_b + (\epsilon - {g^2\over\omega})n_f \eqno(16)$$
with $n_f = 0,1$ and $n_b = 0,1,2,\cdots$. It is clear now that when Eq.(4) is
satisfied the theory is nothing other than the supersymmetric oscillator in
terms of these new variables. The supersymmetric partner states are the
conventional ones in the quanta of the redefined variables. We also note that
when $\epsilon - {g^2\over\omega} < 0$, the ground state of the theory becomes
fermionic [9] whereas if $\epsilon - {g^2\over\omega} = 0$,
the fermions completely
drop out of the theory. Simple as the Hamiltonian in Eq.(3) may appear to be,
it really
has a rich structure. It is clear now (see, e.g., [6-8]) from the form of
the Hamiltonian in
Eq.(14) and the unitary transformation in Eq.(12) that one could also
have started
with a more complicated interacting Hamiltonian in Eq.(3) which would have
resulted in a supersymmetric, interacting Hamiltonian in terms of the
variables $b$ and $f$.

To conclude, we have shown in a simple quantum mechanical model how
supersymmetry can be dynamically generated in the presence of interactions even
when the free theory may not be supersymmetric. It remains to be seen if and
how this idea can be generalized to relativistic quantum field theories. The
properties of such theories would be quite interesting to investigate.

We would like to thank Prof. S. Okubo for comments and discussions.
This work was supported in part by the U.S. Department of Energy Grant No.
DE-FG-02-91ER40685. M.H. would like to thank the
 Funda\c c\~ao de Amparo a Pesquisa
do Estado de S\~ao Paulo for financial support.

\vfill\eject

\noindent {\bf References}

\medskip

\item{1.} See, for example, P. Fayet and S. Ferrara, Phys. Rep. {\bf
32C} (1977) 249; J. Wess and J. Bagger, `` Supersymmetry and Supergravity "
, Princeton Univ. Press, Princeton, NJ 1983; S.J. Gates, M.T. Grisaru, M. Rocek
and W. Siegel, `` Superspace, or One Thousand and One Lessons in
Supersymmetry ",
Benjamin-Cummings, Reading, Mass. 1983; P. van Nieuwenhuizen, Phys. Rep. {\bf
68C} (1981) 264.

\item{2.} M.B. Green, J.H. Schwarz and E. Witten, `` Superstring Theory ",
Cambridge Univ. Press 1987.
\item{3.} F. Iachello and P. van Isacker, `` The Interacting Boson-Fermion
Model ", Cambridge Univ. Press 1991.
\item{4.} B.A. Kupershmidt, `` Elements of Superintegrable System ",
Reidel, Dordrecht 1987.
\item{5.} See, for example, H. Baer et al., `` Low Energy Supersymmetric
Phenomenology ", FSU-HEP-950401, [hep-ph/9503479].
\item{6.} A. Das, `` Field Theory: A Path Integral Approach ",
World Scientific, Singapore 1994.
\item{7.} E. Witten, Nucl. Phys. {\bf B188} (1981) 513.
\item{8.} P. Salomonson and J.W. van Holten, Nucl. Phys. {\bf 196} (1982) 509.
\item{9.} This model was used earlier as a toy model in the study of bound
states in quantum
field theories in S.P. Misra, Indian J. Phys. {\bf 61B} (1987) 287.
\item{10.} See also, A. Das in  `` A Gift for Prophecy ", ed. E.C.G.
Sudarshan, World Scientific, Singapore 1995.
\item{11.} R.J. Glauber, Phys. Rev. Lett. {\bf 10} (1963) 84;
E.C.G. Sudarshan, Phys. Rev. Lett. {\bf 10} (1963) 277.
\end